\documentclass[preprint,12pt]{elsarticle}

\usepackage{graphicx}
\usepackage{miller}
\usepackage{fullpage}
\usepackage{dcolumn}
\usepackage{bm}
\graphicspath{figures}
\renewcommand{\selectlanguage}[1]{}

\begin{document}

\title{Sensitivity of Multislice Electron Ptychography to Point Defects: A Case Study in SiC}

\author{Aaditya Bhat}
\author{Colin Gilgenbach}
\author{Junghwa Kim}
\author{Michael Xu}
\author{Menglin Zhu}
\author{James M.~LeBeau}
\ead{lebeau@mit.edu}

\affiliation{%
Department of Materials Science and Engineering, Massachusetts Institute of Technology, Cambridge, MA, 02139 United States
}%

\date{\today}

\begin{abstract}

Here, we evaluate multislice electron ptychography as a tool to carry out depth-resolved atomic resolution characterization of point defects, using silicon carbide as a case study. Through multislice electron scattering simulations and multislice ptychographic reconstructions, we investigate the phase contrast arising from individual silicon vacancies, antisite defects, and a wide range of substitutional transition metal dopants (V\textsubscript{Si} to W\textsubscript{Si}) and potential detectability. Simulating defect types, positions, and microscope conditions, we show that isolated point defects can be located within a unit cell along the sample's depth. The influence of electron energy, dose, defocus, and convergence semi-angle is also explored to determine their role in governing defect contrast. These results guide experiments aiming to analyze point defects with multislice electron ptychography.

\end{abstract}

\maketitle

\section{Introduction}

Analyzing the structure of point defects is crucial for gaining insights into the structural properties of many engineered materials, from semiconductors \cite{McCluskey2020-oe} to oxides \cite{Hughes1972-td}. As an example, color center defects in silicon carbide (SiC) can possess properties suitable for optical and quantum applications, for example, as single photon emitters in the telecom range \cite{Castelletto2020-ro}. These color centers arise from a rich class of single point defects \cite{Harmon2022-qz} and defect complexes \cite{Christle2015-pc}, both intrinsic and extrinsic \cite{Harmon2022-qz,Castelletto2020-ro,Bockstedte2003-wb}. Recently, divacancies (vac\textsubscript{Si}-vac\textsubscript{C}) and substitutional transition metal dopants (TM\textsubscript{Si}) complexed with carbon vacancies have been shown to form single photon emitters \cite{Christle2015-pc, Wolfowicz2021-sd, Anderson2022-sf}. However, quantifying these point defects directly remains challenging, particularly when interacting with extended defects and interfaces.

In electron microscopy, high-angle annular dark field (HAADF) STEM  has been used to identify point defects in materials, including substitutional dopants \cite{Voyles2003-uc, Ishikawa2020-ux, Saito2017-fs, HL_Xin2007-td}, vacancies \cite{Jinwoo_Hwang2013-bv, Kim2016-tc, Feng2015-zm}, and interstitials \cite{Johnson2019-ng, Kim2023-nl}. There are, however, a number of limitations. First, the sample is required to be very thin (typically 10 nm), which limits the volume probed and thus requires materials with high defect point concentrations. This is true even for through-focal series HAADF imaging, which extends the spatial mapping of point defects to 3D as demonstrated for heavy dopants in amorphous \cite{Klaus_van_Benthem2006-ag} and crystalline \cite{Ishikawa2020-ux, Saito2017-fs} materials. Second, very clean surfaces and precise sample orientation are needed to maximize defect contrast \cite{Jacob_T_Held2017-xx, Mittal2011-tv}. As a consequence, when the sample is not perfectly flat, defect concentrations are low (below ~$10^{19}$ cm$^{-3}$ or 0.1 at\%), and/or when there is limited atomic number contrast between defect and host, such investigations remain exceedingly challenging \cite{Propst2023-ff, Mittal2011-tv}.  

In contrast to prior direct imaging, computational phase retrieval approaches \cite{Miao2025-rd} enable new opportunities for point defect detection and quantification. For example, multislice electron ptychography uses a 4D dataset of diffraction patterns acquired with a scanned, often defocused, probe to iteratively reconstruct even relatively thick samples \cite{Chen2021-ep, Maiden2012-xg, Gao2017-pz}. This is achieved by accounting for dynamical diffraction through the multislice forward model that recovers depth-resolved information  where surface features can be well separated from the interior \cite{Karapetyan2024-bd}. Moreover, because the MEP reconstructs the incident probe and collects the majority of electrons that passed through the sample, the ultimate resolution is  limited only by thermal vibrations \cite{Chen2021-ep} and significantly improves the signal to noise ratio for light element detection \cite{LiImagingwithptycho}. 

Recent advances in MEP indicate that single point defect detection and quantification can be achieved with $\sim\!3$ nm depth resolution in experiments, with current instrumentation and in sample thicknesses up to 30 nm \cite{Chen2021-ti,Dong2024-io,Dong2025-uu}. For example, imaging of heavy dopants \cite{Dong2025-uu} or interstitials \cite{Chen2021-ti}, and even oxygen vacancies \cite{Dong2024-io} have been demonstrated. While the possibilities for robust identification of point defects using ptychography are promising, the technique's sensitivity to subtle differences in the projected potential and noise must be established in the context of defect types, geometry, and acquisition conditions.

In this article, we explore the sensitivity of multislice electron ptychography to the detection of substitutional point defects and complexes via 4D STEM simulations and subsequent MEP reconstructions. As a case study material, 4H-SiC is selected as a technologically relevant material that helps to illuminate the prospects and challenges of MEP for point defect quantification. Various candidate defect structures are considered across different microscope conditions, including electron energy, convergence semi-angle, electron dose, and acquisition parameters. At each condition, the sensitivity of MEP to defect type and position are evaluated and quantified. These results help to establish the boundaries of point defect characterization with MEP and enhance sensitivity to particular defects.

\section{Methods}

\subsection{4D STEM Simulations}

\begin{figure}[ht]
    \centering
    \includegraphics[width=3.1in]{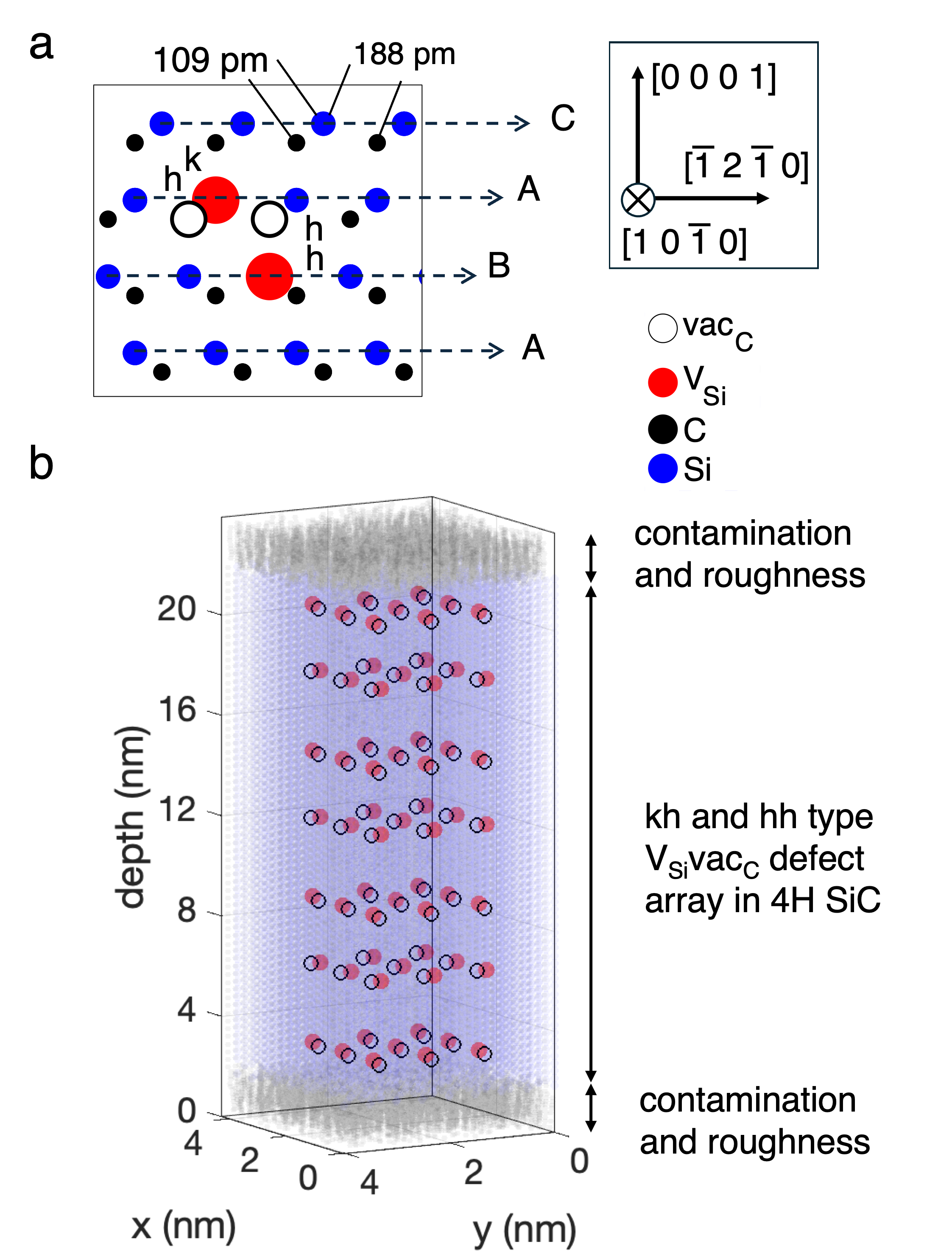}
    \caption{(a) Schematic of the $kh$ and $hh$ complex configurations viewed along  \hkl[10-10] of 4H SiC which consists of alternating Si and C planes with ABAC stacking. (b) Schematic representation of a supercell containing an array of TM\textsubscript{Si}-vac\textsubscript{C} defect complexes, with surface roughness/carbon contamination present. }
    \label{fig:array}
\end{figure}

A supercell of 4H-SiC was constructed with dimensions 4.3 nm $\times$ 4.0 nm $\times$ 24.0 nm to contain the electron probe within the simulation volume completely. The sample thickness was chosen as it can be readily achieved through either focused ion beam preparation or conventional polishing and ion milling \cite{Voyles2003-uc, Chen2024-va}. The supercell was oriented with the electron beam along \hkl[10-10]. Along this orientation, Si and C exhibited ABAC stacking with $h$ (hexagonal) and $k$ (cubic) sites,  indicated in Figure \ref{fig:array}a, that can be readily distinguished. A 2 nm thick amorphous carbon layer was added to the entrance and exit surfaces to approximate surface contamination \cite{Ricolleau2013-tk, Hugenschmidt2023-pj}, Figure \ref{fig:array}b,. 

The simulated defect-containing supercells consisted of arrays of $kh$ and $hh$ type defect complexes, as shown in Figure \ref{fig:array}b. Varying the depth of a Si site defect, X\textsubscript{Si}, complexed  with a carbon vacancy, vac\textsubscript{C}, provided a test of the depth sensitivity and variation in the reconstructed phase at different positions. X was varied to include vacancies ($Z = 0$), carbon antisites (C\textsubscript{Si}), and substitutional dopants with $Z$ between 23 and 74.  For simulations investigating depth sensitivity and microscope parameter dependence, substitutional vanadium defects, V\textsubscript{Si}, were considered in complexes as they are an established spin active defect in SiC \cite{Wolfowicz2021-sd} with a relatively low atomic number difference with Si ($\Delta Z = 9$), thus providing a relevant test case for sensitivity.

Frozen lattice multislice simulations  were used to generate 4D STEM datasets using a custom Python package based on, and validated against, programs developed by E.~Kirkland \cite{Kirkland-ac}. Unless specified otherwise, simulations used a scan step of 50 pm, electron energy of 200 keV, 10 nm of probe overfocus, and a convergence semi-angle of 25 mrad. The detector considered was 128$\times$128 pixels with a diffraction pixel size of 1.0 mrad. Thermal vibrations were incorporated using root mean square (rms) thermal displacements of 5.3 and 5.4 pm for Si and C \cite{Reid1983-it}, respectively.  Using the Einstein approximation, 50 thermal configurations were incoherently averaged. The amorphous carbon layer was assigned rms thermal displacements of approximately 20 pm ~\cite{Mullner1979-yq}

The effect of the detector point spread function (PSF) was incorporated using that of an Electron Microscope Pixel Array Detector (EMPAD) \cite{Tate2016-uu}, for an electron beam energy matching the 4D STEM simulations. Unless otherwise noted, shot noise was added at each detector pixel, assuming a beam current of 7.5 pA and a dwell time of 1 ms. For the scan parameters above, this corresponded to a total simulated dose of $1.8 \times 10^7 \;\mathrm{e^-/nm^2}$.  

\subsection{Reconstructions}

The fold\_slice \cite{Chen2021-ep,Jiang_undated-ko} fork of Ptychoshelves \cite{Wakonig2020-kg} was used for multislice ptychographic reconstructions. The least squares maximum likelihood iterative engine with GPU-acceleration \cite{Thibault2012-eg,Thibault2013-or,Tsai2016-uc,Odstrcil2018-ud} was used. Mixed-state reconstructions \cite{Chen2020-ou} with eight probe modes were utilized to account for probe partial coherence in all cases. The reconstructions included a total object thickness of 24 nm composed of 1 nm-thick slices. 

Based on the 4D STEM simulation parameters above, the reconstructed objects had a pixel size of 20 pm. For simulations with electron energy below 200 keV, the diffraction patterns were zero-padded before reconstruction to match the real space sampling. For an electron energy of 300 keV, the diffraction patterns were not cropped to prevent loss of scattered signal, which resulted in reconstructions with object pixel sizes of 15.6 pm. Since all reconstructions were interpolated using smoothing splines prior to data analysis, this additional sampling is not expected to influence contrast at defect sites \cite{Reinsch1967-uq, Silverman1985-em}. The reconstructed phase background \cite{Fannjiang2019-jc} was removed by fitting a plane to the image \cite{Zavodov2021-hg}. Moreover, due to the added carbon contamination at the surfaces and a $\sim$2 nm depth resolution in simulations, a diffuse inhomogeneous background in slices near the surface \cite{Karapetyan2024-bd} was removed by cropping the top and bottom 4 slices (4 nm from each surface) of the reconstruction.  

\section{Results and Discussion}

\subsection{Reconstructed Phase Evaluation} 

\begin{figure}[ht]
    \centering
    \includegraphics[width=3.1in]{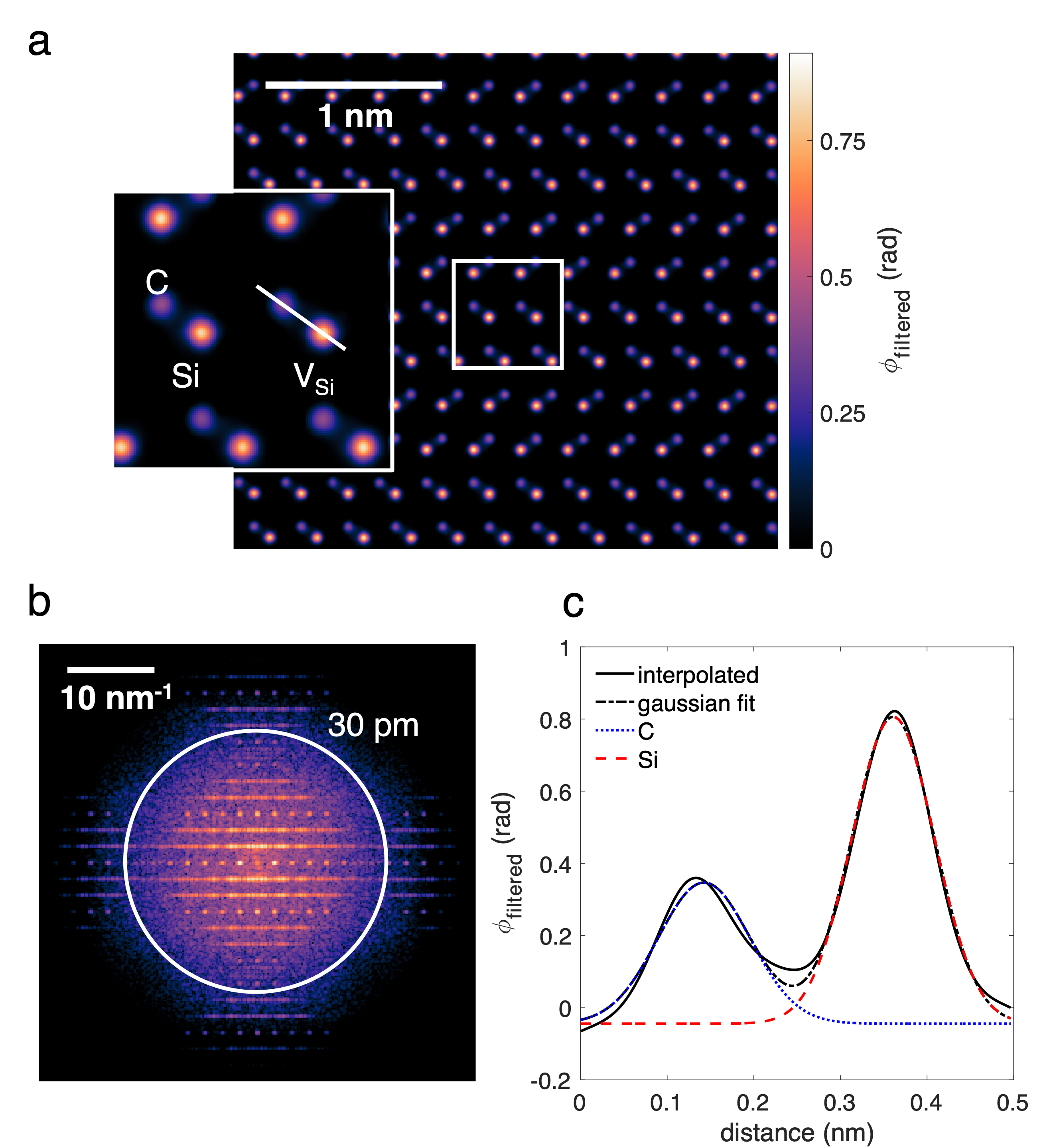}
    \caption{(a) Example reconstructed object slice (resampled using cubic spline interpolation) showing resolved Si and C atom columns and V\textsubscript{Si} defect annotated in inset. (b) $<$30 pm resolution as seen from the Fourier transform of the slice with a Hann filter applied. (c) Line profile across the near neighbor V\textsubscript{Si} C pair showing well resolved phase with Gaussian fitting, but inaccurate fit to the reconstructed object phase.}
    
    \label{fig:res}
\end{figure}

Ptychographic reconstructions of the simulated 4D STEM data  enable clear identification of Si and C sublattices via atomic number-dependent phase contrast, as shown in inset Figure \ref{fig:res}a. Furthermore, the lateral information limit is less than 30 pm as seen in Figure \ref{fig:res}b, which is more than sufficient to separate the Si-C nearest neighbor pairs. A Gaussian pair fit to the interpolated Si and C sites in Figure \ref{fig:res}c show that contributions of phase from nearest neighbor sites to the peak phase of a given site is negligible. The fit peak phase at both Si and C sites shows 1.9\% and 3.6\% difference respectively from the underlying interpolated peak phase due to inherent asymmetry in the distributions at both sites. Although small relative to the absolute phase, these inaccuracies are significant when compared to phase contrast due to substitutional point defects discussed below. The interpolated peak phase at the well-resolved Si and C sites, denoted $\phi$\textsubscript{Si} and $\phi$\textsubscript{C}, thus provides a more robust approach to quantify contrast.

\begin{figure}[ht]
    \centering
    \includegraphics[width=3.1in]{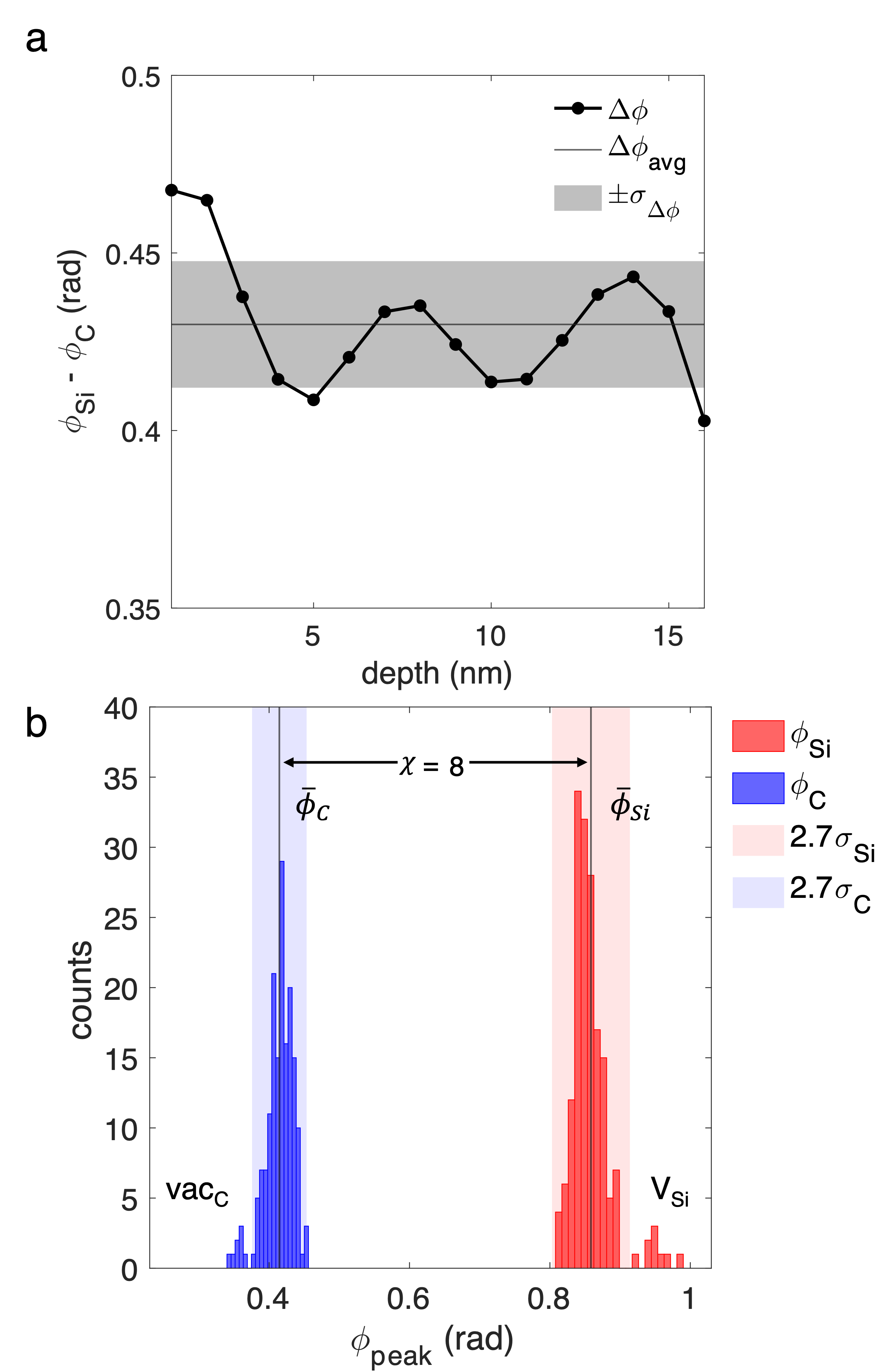}
    \caption{(a) Variation of difference in mean Si and C atom column phase through depth showing non-uniformity and $>$10\% change. (b) Collective histogram of peak Si and C site phases in a slice containing an array of hh type V\textsubscript{Si} vac\textsubscript{C} defects with $\pm$2.7$\sigma_{site}$ ($\chi$=1) and defects with phase beyond these limits highlighted.}   
    
    \label{fig:datan}
\end{figure}

Slice-dependent offsets \cite{Fannjiang2019-jc} and scaling present in reconstructions, seen in the depth dependent variation of mean Si can C site phase difference in Figure \ref{fig:datan}a, requires additional normalization to enable meaningful comparisons across slices. This variation is more than 10\% of the difference between $\overline{\phi_\mathrm{Si}}$ and $\overline{\phi_\mathrm{C}}$ in some slices and larger than the phase difference due to some of the substitutional dopants considered as discussed further below. Furthermore, at defect free Si and C sites, variations in the phase distribution are observed at the Si and C sites even under identical reconstruction conditions (Figure \ref{fig:datan}b), which are present in reconstructions with finite iterations, irrespective of conditions of infinite dose.

To ensure measurements from the slices are independent of offsets and scaling, the phase difference relative to the mean phase of each element is analyzed, which is then normalized by the mean phase difference between Si and C elements, $\overline{\phi_\mathrm{Si}-\phi_\mathrm{C}}$,within each slice. This `site-contrast', $\chi$, defined as follows, captures the relative scattering strengths of Si, C and dopants:

\begin{equation}
    \chi_{\mathrm{site}} = \frac{\phi_{\mathrm{site}} - \overline{\phi}_{\mathrm{site}}}{(\overline{\phi}_{Si}-\overline{\phi}_{C})/\Delta Z}
\end{equation}

\noindent where $\overline{\phi}_{Si,C}$ denotes the mean phase value of the respective element (Si or C). The $\Delta Z$ in the denominator normalizes the phase difference to the difference in atomic numbers of Si and C, here $\Delta Z = 8$, to obtain a material system independent metric for dopant contrast. This ensures consistency across datasets and slices since scattering at defect sites relative to this difference is independent of any offsets or scaling, and nominally only depends on relative scattering strengths. Furthermore, normalizing for scaling factors results in a metric that depends solely on the reconstructed potential and is hence independent of the electron accelerating voltage dependent interaction parameter.

The distributions of Si and C site contrast, shown in Figure \ref{fig:datan}b, have standard deviations ($\sigma$) of 0.36 and 0.33, respectively, providing baselines to assess the detectability of point defects. Assuming a Gaussian distribution for defect free sites, atom columns with site contrast beyond $\chi$=$\pm$1 ($\pm$2.7$\sigma$) are identifiable as defects with $\approx$99.5\% true positive detection rate here. Although detection of low $|\Delta Z|$ defects is limited by variations at defect free sites, the mean site contrast is primarily dependent on scattering strength of defects relative to silicon and carbon. Hence, site contrast provides a relative-scattering-strength-based metric is expected to be robust against variations resulting from sample quality and experimental conditions and is thus used in the following analysis.

\subsection{Depth and phase variation at point defects}

\begin{figure}[ht]
    \centering
    \includegraphics[width=3.1in]{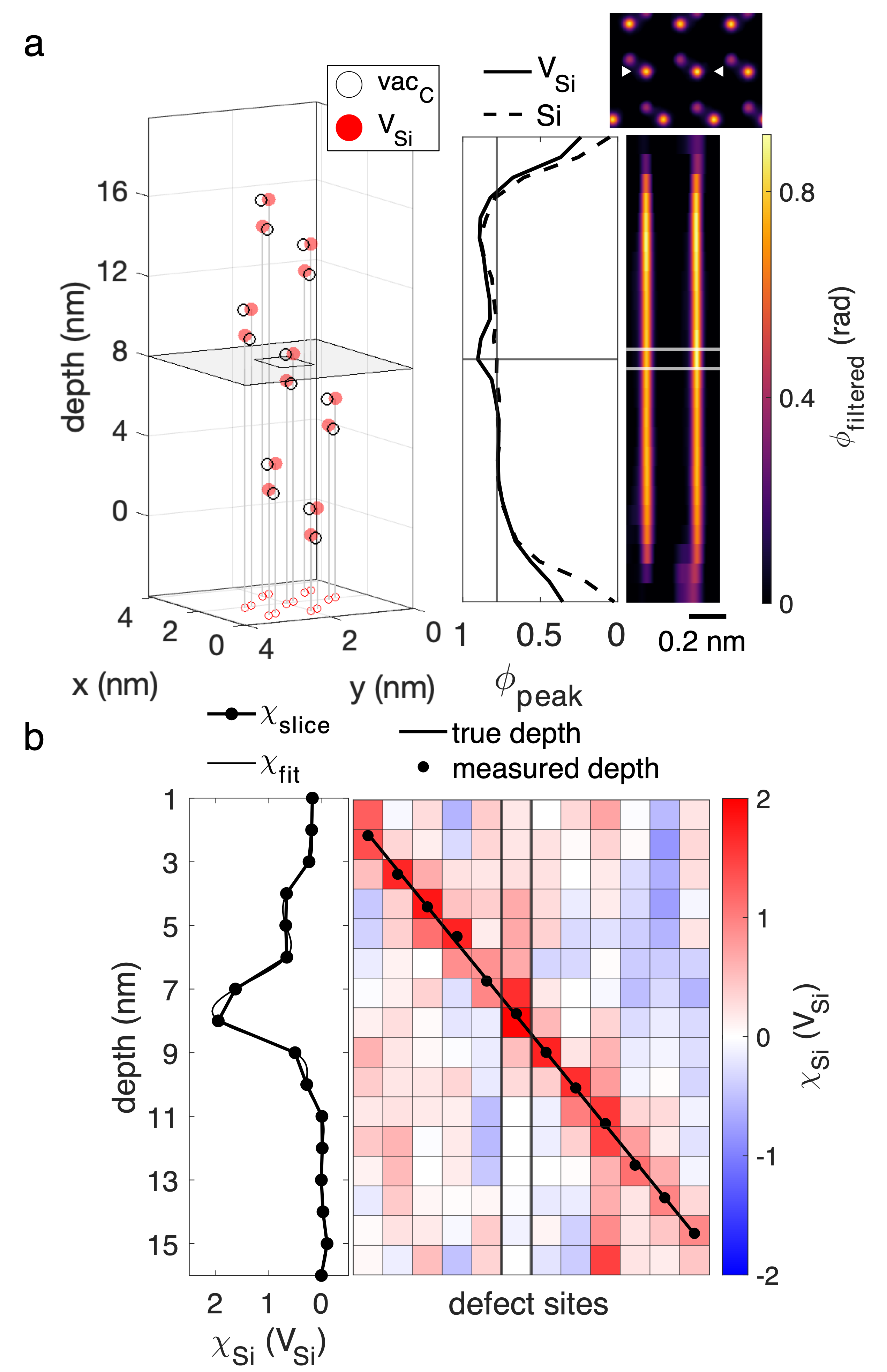}
    \caption{a) Schematic of the array of V\textsubscript{Si} vac\textsubscript{C} defects at various depths and an example cross section of the reconstructed multislice phase object through depth and across a line containing neighboring Si columns with and without a vanadium substitution. (b) Site contrast variation through depth at defect sites showing the true position in the structure compared to measured position in depth. Measured position is estimated using smoothing cubic spline interpolation with an example fit shown at the highlighted defect column.}
    \label{fig:depth}
\end{figure}

The capability of MEP to localize point defects in 3D is evaluated by examining site contrast for columns containing vanadium substitutions (V\textsubscript{Si}, $\Delta Z$=9) at different depths, as illustrated in Figure \ref{fig:depth}a.  Regardless of their position along the electron beam direction, V\textsubscript{Si} defects exhibit increased site contrast, 1.14$\pm$0.36,  compared to those of their neighboring defect-free Si sites. As a function of depth, the site contrast increases in the layers above and below the defect location. Further, by fitting a cubic smoothing spline to the phase as a function of depth, the depth resolution is approximately 1.9 nm as determined from the full-width half-at-half-maximum, Figure \ref{fig:depth}b. Based on the standard error, the  depth precision is 0.1 nm, or well within a single unit cell along the beam direction. Combined with the above power law fit, these results indicate that MEP can provide atomic number contrast down to the level of point defects, even those 'buried' within a sample. 

\begin{figure}[ht]
    \centering
    \includegraphics[width=3.1in]{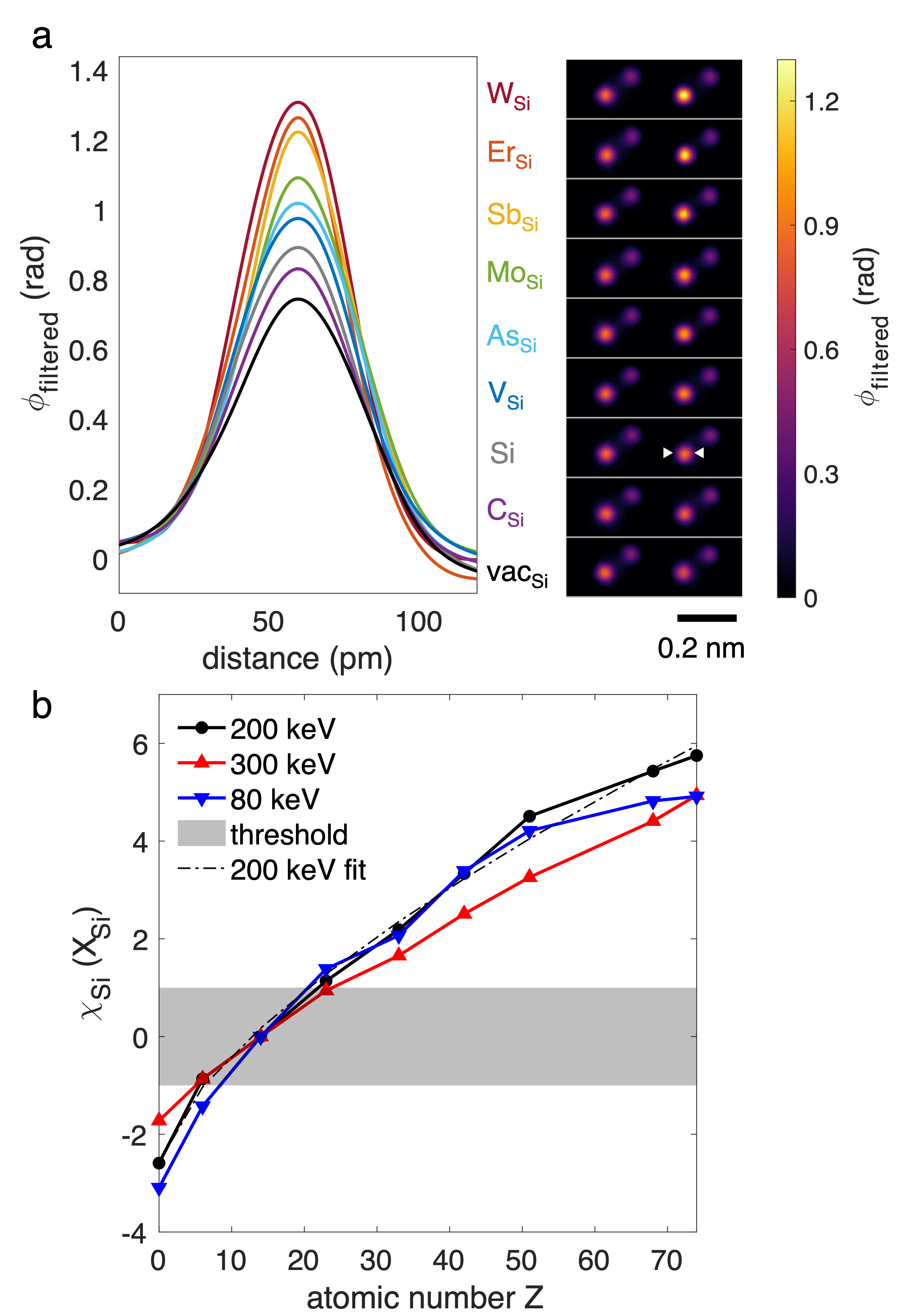}
    \caption{(a) Line profiles across defect sites containing intrinsic and extrinsic substitutions show increasing peak phase with atomic number and (b) variation of atomic number dependence of site contrast with incident electron energy. Z = 0 corresponds to vacancies.}
    \label{fig:atnum}
\end{figure}

To investigate site contrast changes due the the presence of point defects,  metal substitutions at Si sites (TM\textsubscript{Si}) and vacancies (vac\textsubscript{C}) are considered.  Heavy substitutions with large atomic number differences ($\Delta Z$) from Si, e.g.~molybdenum (Mo\textsubscript{Si}, $\Delta Z$=28) and erbium (Er\textsubscript{Si}, $\Delta Z$=54), exhibit significantly increased reconstructed phase as shown in Figure \ref{fig:atnum}a. The corresponding site contrasts are 3.34 (Mo\textsubscript{Si}) and 5.43 (Er\textsubscript{Si}), which deviate significantly from the pristine Si-C site contrast and makes them visually discernible. Silicon vacancies (vac\textsubscript{Si}, $Z=0$), on the other hand, yield a site contrast is -2.6, or ~7$\sigma$ below pristine Si sites. 

Across a range of possible dopants and defects, the site contrast for even single defects follows a $Z^{0.68}$ relationship,  as shown in Figure \ref{fig:atnum}b. This is  consistent with the relationship of defect free crystals in Ref.~\cite{Chen2021-ep}. In principle then, analyzing the reconstructed phase  offers a possible method to identify the specific type of point defect. It is important to note, however, that identifying point defects in the low $|\Delta Z|$ regime remains challenging. In particular for low $|\Delta Z|$ defects, site contrast often fall within the $|\chi_{Si}(X_{Si})|<1$ range, as indicated by the gray highlighted region
in Figure \ref{fig:atnum}b. Carbon antisites (C\textsubscript{Si}, $\Delta Z$=-8), for example, exhibit a site contrast of -0.87 ($\sigma$ std), which lies within $0>\chi_{Si}(X_{Si})>-1$, making it difficult to distinguish subtle variations at defect free sites from true defect-induced contributions to the phase shifts. 

\subsection{Improving site contrast}

To improve the detectability of point defects, the site contrast should be maximized compared to the distribution of defect-free Si columns. The following investigates how site contrast varies under different incident electron energy, dose, and other acquisition parameters because they can have a significantly affect reconstruction quality and $\Delta Z$ sensitivity \cite{Gilgenbach2024-ug}. V\textsubscript{Si} and vac\textsubscript{C} point defects serve as benchmarks since $\chi_{Si}(V_{Si})\sim1$, which is  the lowest detectable $\Delta Z$ point defects considered here. Since defect dynamics relevant to color centers are predominantly driven by Si site substitutional defects \cite{Bockstedte2003-wb, Lee2021-me, Zhang2023-ub}, they are the primary defects considered here. 

\subsubsection{Incident Electron Energy}   

Since the strength of electron scattering is influenced by incident electron energy, it is expected to impact defect site contrast. To investigate this effect, three commonly used electron energies – 80 keV, 200 keV, and 300 keV are compared in Figure \ref{fig:atnum}b. Site contrast for V\textsubscript{Si} defects is 1.38 at 80 keV, representing a 21\% increase compared to 200 keV. While site contrast is enhanced for small $\Delta Z$, the improvement is reduced for higher $\Delta Z$ ($\Delta Z > 25$). 

Carbon vacancies (vac\textsubscript{C}), one of the most abundant defects in SiC \cite{Bockstedte2003-wb}, are challenging to image given their low $\Delta Z$= -6 that results in a  low site contrast ($\chi_C(vac_C)=0.93$) . For these defects, their site contrast is enhanced to -1.44 at 80 keV, which is a 55 \% increase compared to that at 200 keV, which can result in a higher detection rate. This behavior is also observed for C\textsubscript{Si}, where site contrast improves from -0.87 at 200 keV to -1.43 at 80 keV. Increasing the electron energy to 300 keV, on the other hand, reduces the site contrast to 0.94 -- a 17 \% decrease compared to 200 keV. 

Identifying point defects is more difficult at  300 keV for all substitutions, despite lower wavelength and hence higher resolution. In contrast, 80 keV is best suited for detecting light element substitutions (low $\Delta Z$ defects) or vacancies.  The non-monotonic trend relating site contrast to energy might be attributed to the pronounced influence of dynamical diffraction effects at low energies. The increased interaction parameter at lower keV results in stronger multiple scattering for the same projected atomic potential \cite{Reimer2008-nw, Kirkland-ac}, and hence, cannot be represented equivalently in reconstructed object slices of equal thickness. Thus, site contrast is expected to vary with electron accelerating energies despite the relative scattering strengths of atoms being identical.

Although  higher defect contrast is obtained at 80 keV, obtaining converged reconstructions requires additional iterations due to padded datasets necessary to obtain object sampling equivalent to that at 300 keV with higher probe convergence semi-angles for equivalent extent of the probe in reciprocal space. Reconstructions for simulations at 60 keV fail with comparable parameters. Additionally, the increased inelastic to elastic cross section ratio at lower energies can increase incoherence in diffraction patterns, resulting in further challenges in reconstructions \cite{Dong2024-io}.

Damage is another important consideration. Because the knock-on threshold for light elements is lower \cite{Egerton2010-db}, e.g.~ the threshold for carbon vacancy generation has been reported to be $<$150 keV \cite{von-Bardeleben2000-rt, Knezevic2022-uy}, 300 keV electrons have been used to create both silicon and carbon vacancies in SiC  \cite{von-Bardeleben2000-rt, Beyer2011-nt}. This suggests that lower accelerating energies, despite challenges in achieving successful reconstructions, may be preferred for point defect detection. On the other hand, radiolysis damage increases at lower accelerating energies in some materials, potentially complicating this analysis \cite{Egerton2019-sv}. Overall, point defect classification in materials containing light elements is expected to be considerably easier at low energy, given higher site contrast and reduced damage.

\subsubsection{Dependence on dose}

Dose plays a dual role in the reliable quantification of defects, through increased noise in the acquired data at low doses and structural changes induced in the sample by incident high energy electrons at high doses. To investigate the role of dose on site contrast, the simulated dose is varied from 0.6--$75 \times 10^6 \;\mathrm{e/nm^2}$ in arrays of V\textsubscript{Si} vac\textsubscript{C} defects described above. 

\begin{figure}[ht]
    \centering
    \includegraphics[width=3.1in]{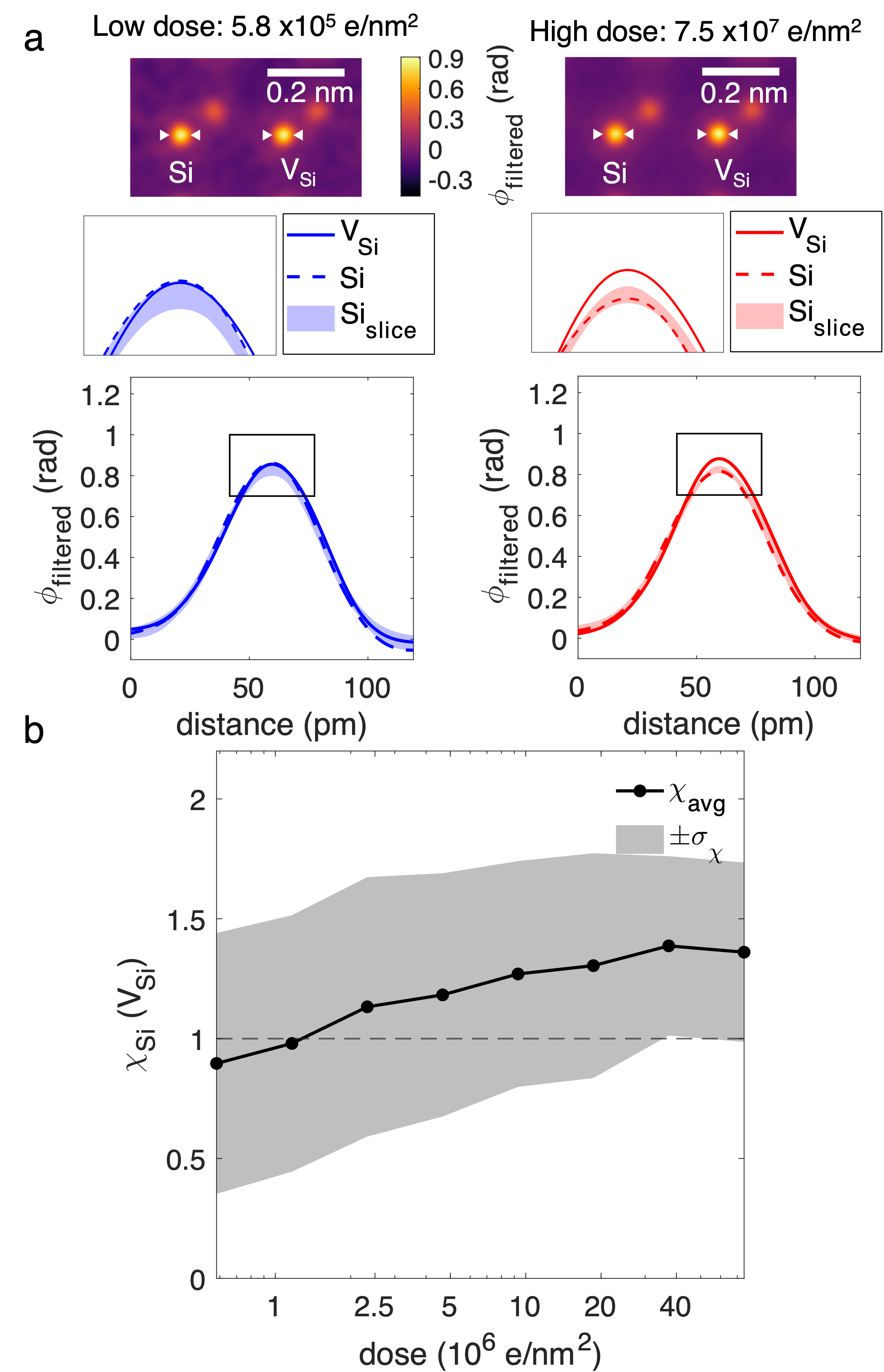}
    \caption{(a) Reconstructed phase and line profiles across neighboring Si sites with and without V\textsubscript{Si} defects at a low ($0.6 \times 10^6 \;\mathrm{e/nm^2}$) and high dose ($75 \times 10^6 \;\mathrm{e/nm^2}$). Variation of phase across Si sites in the slice is shown for reference. (b) Mean site contrast at defect sites decreases in magnitude sharply at low doses as seen in the variation of site contrast with dose (dose shown with a log scale).}
    \label{fig:dose}
\end{figure}

The variation of site contrast with dose in Figure \ref{fig:dose} shows that at doses above $1.2 \times 10^6 \;\mathrm{e/nm^2}$, mean site contrast shows a small change of $7\%$ (from 1.27 to 1.36) approaching a dose of $75 \times 10^6 \;\mathrm{e/nm^2}$. Below this lower limit, reconstructions show distinct Si and C atom columns, but the standard deviation of defect site contrast increases above 0.5, and $>10\%$ of defects are not identifiable. 

At the lowest dose considered, point defect detection is not possible for vanadium due to the site contrast at V\textsubscript{Si} being lower than the variation in Si site phase as seen in line profiles of the filtered reconstructed phase in Figure \ref{fig:dose}. Moreover, the reconstructed slices at low doses exhibit increased background noise that is absent at high doses, as shown in Figure \ref{fig:dose}. This noise reflects a poorer quality of reconstruction in addition to significantly reduced defect site contrast associated with decreasing dose. 

Additionally, Poisson noise at the detector is known to affect reconstruction quality  \cite{Jiang2018-yl}. At scan step of 50 pm and a dwell time of 1 ms, such as is typical with an EMPAD, the tested doses correspond to a beam current ranging from 0.25 pA to 30 pA. Each diffraction pattern receives approximately $1.5 \times 10^3$ to $2 \times 10^5$ incident electrons, and with the detector parameters used here, this results in each pixel in the bright field disk receiving an average of $<$4 electrons at the lowest dose considered for $\sim 30 \%$ noise to signal ($1/\sqrt{n}$). Poisson noise thus imposes a lower limit on dose and quantitative recovery of the object phase from the diffraction patterns.

While higher doses per diffraction pattern are preferred for enhanced signal-to-noise, the possibility for accompanying beam induced damage to the sample from the entire scan must be considered. For atomic resolution multislice electron ptychography, in particular, smaller scan step sizes are needed to achieve high quality reconstructions compared with single-slice methods \cite{Gilgenbach2024-ug, Chen2021-ti, Leidl2024-qj, D-Alfonso2016-az}. This translates to higher doses in experiment, which will increase the likelihood of beam induced defect generation/migration as well as structural transformations \cite{Ishimaru2003-li, Kerbiriou2007-sg, Mishra2017-qz, Ishikawa2014-nb}. 

\subsubsection{Acquisition parameters}

\begin{figure}[ht!]
    \centering
    \includegraphics[width=3.1in]{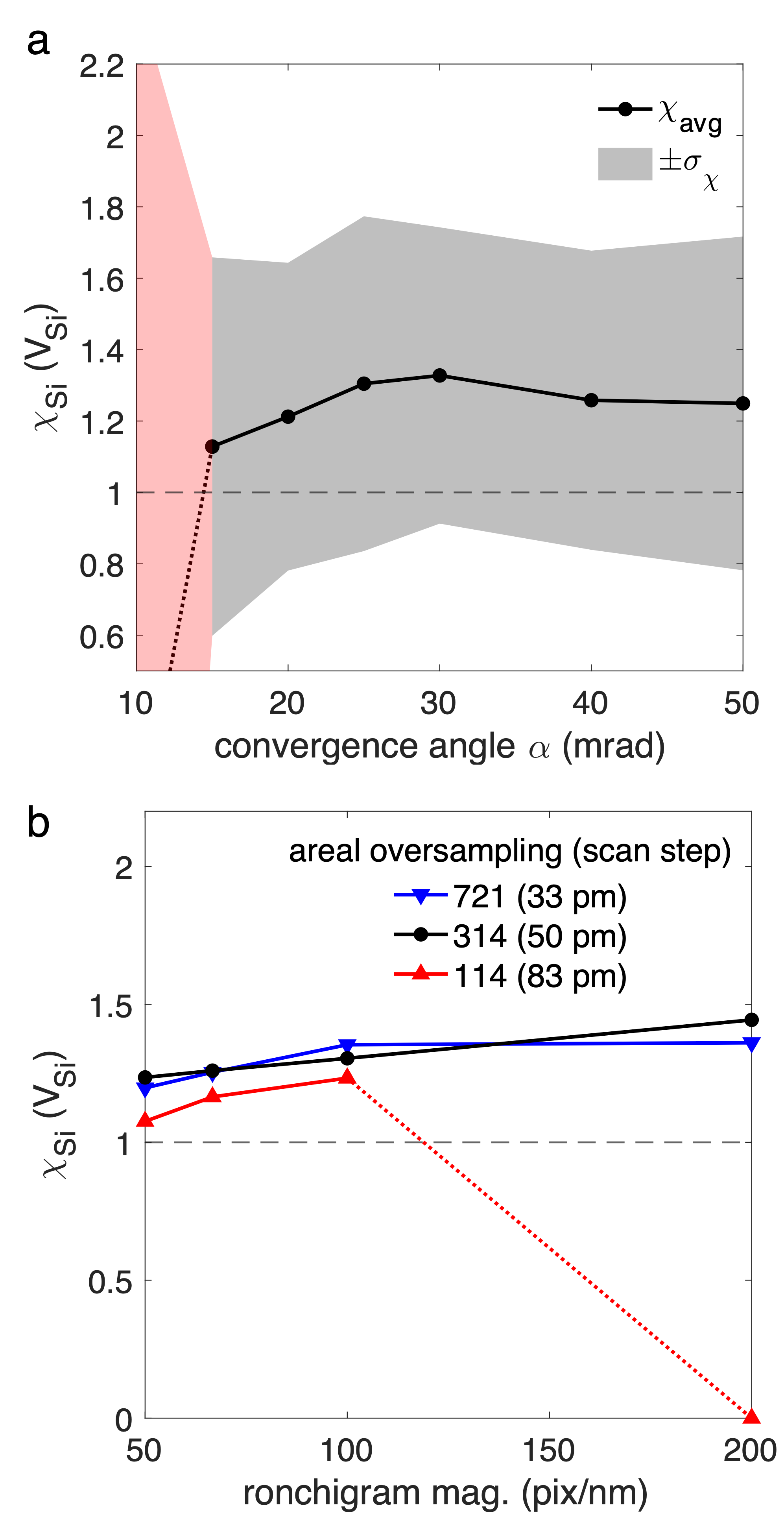}
    \caption{(a) Variation of mean V\textsubscript{Si} site contrast with convergence semi-angle. Sharp decrease in mean site contrast and increase in standard deviation is seen with a probe convergence semi-angle of 10 mrad. (b) Variation of mean V\textsubscript{Si} site contrast with Ronchigram magnification shows negligible change with scan step except at high scan steps and high Ronchigram magnification where reconstructions fail due to low area oversampling.}
    \label{fig:acqpar}
\end{figure}

In addition to the parameters considered above, defocus, scan step, and probe convergence semi-angle influence quality and contrast in the the reconstructions for a given dose ($1.8 \times 10^7 \;\mathrm{e/nm^2}$) and electron energy (200 keV). Independent variation of the probe convergence semi-angle shows the optimum for site contrast lies between 25 and 30 mrad as seen in Figure \ref{fig:acqpar}a. Site contrast for V\textsubscript{Si} increases by 10 \% from 1.21 at 20 mrad to 1.33 at 30 mrad. For a convergence semi-angle of 10 mrad, reconstructions fail with an equivalent set of parameters. This suggests convergence semi-angle has limited influence on site contrast beyond general requirements for successful reconstructions \cite{Gilgenbach2024-ug}. 

The optimum defocus and scan step size are not independent due to their coupled effect on contrast at defect sites. Instead, dimensionless metrics derived from these parameters can be considered. Specifically, the Ronchigram magnification, $\frac{1}{\lambda k_dd}$ where k\textsubscript{d} is the angular diffraction pixel size and $d$ is defocus, and areal oversampling, $\frac{\pi (\alpha d)^2}{s^2}$ where $\alpha$ is the convergence semi-angle and $s$ is the scan step, are more directly correlated to the quality of reconstructions\cite{Gilgenbach2024-ug}. Areal oversampling is a feature of the scan and probe geometry alone, and decoupled from the object, whereas Ronchigram magnification quantifies relative length scales of features in the object and the probe, and hence can be considered independently.

Decreasing overfocus increases the Ronchigram magnification, which in turn produces an increased sensitivity to point defects. Specifically, the mean site contrast for V\textsubscript{Si} increases by 16\% (from 1.24 to 1.44) as overfocus is decreased from 20 nm to 5 nm (Ronchigram magnifications of 50 to 200 pixels/nm) as seen in Figure \ref{fig:acqpar}b. In general, higher Ronchigram magnification increases site contrast as seen in Figure \ref{fig:acqpar}b, provided areal oversampling is sufficient to achieve reconstruction convergence .

Varying areal oversampling, on the other hand, yields small changes in mean site contrast for converged reconstructions. At a large scan step of 83 pm (corresponding to a low areal oversampling) and for a high Ronchigram magnification of 200 pixels/nm, areal oversampling is insufficient for the reconstruction to converge with the fixed set of reconstruction parameters used. A sharp reduction in site contrast, as seen in Figure \ref{fig:acqpar}b, is possible at higher Ronchigram magnifications. This can be attributed to reduced redundancy in the dataset and consequently reduced robustness against noise. While sufficient areal oversampling is critical for reconstruction convergence \cite{Gilgenbach2024-ug}, it has a negligible influence on defect contrast. 

\section{Conclusion}

Depth resolved identification of a wide variety of intrinsic and extrinsic point defects in SiC is possible using multislice electron ptychography, provided suitable microscope and acquisition parameters are chosen. Given optimal conditions, silicon vacancies and substitutions as light as vanadium may be detectable. The estimated depth precision of $\sim\! 0.1$ nm for defects complexes is sufficient to separate them from isolated defects and provide 3D spatial distributions. Among the acquisition parameters investigated, defect site contrast is enhanced when electron energy is reduced, which is also beneficial in reducing beam-induced damage and identification of intrinsic defects such as antisites and carbon vacancies. Moreover, at low doses, the increase in Poisson noise causes a sharp drop in site contrast of defects,  limiting point defect detection in materials that not robust to electron irradiation. Probe convergence semi-angle, defocus and scan step have a relatively limited influence on contrast above a threshold for effective reconstructions. Specifically, a defocus of 5-20 nm, a scan step of $<$50 pm, and a probe convergence semi-angle of 20-30 mrad are target parameters for the purpose of point defect detection. Overall, MEP provides a compelling avenue to resolve structures of point defects, and address the longstanding challenge of robust single substitution detection beyond extremely thin samples with heavy dopants often required by conventional STEM imaging methods. 

\section*{Acknowledgments}

The authors acknowledge support from the Air Force Office of Scientific Research (FA9550-22-1-0370). Data processing was carried out using a modified version of the fold\_slice fork of the cSAXS ptychography MATLAB package developed by the Science IT and the coherent X-ray scattering (CXS) groups, Paul Scherrer Institut, Switzerland. The authors acknowledge the MIT SuperCloud and Lincoln Laboratory Supercomputing Center for providing HPC resources that have contributed to the research results reported within this paper/report.

\bibliographystyle{elsarticle-num}
\bibliography{paperpile}

\end{document}